# Assessing Concordance between RNA-Seq and NanoString Technologies in Ebola-Infected Nonhuman Primates Using Machine Learning

Mostafa Rezapour[1*]; Aarthi Narayanan[2]; Wyatt H. Mowery[3]; Metin Nafi Gurcan[1]

1. Center for Artificial Intelligence Research, Wake Forest University School of Medicine, Winston-Salem, NC 27101, USA
2. Department of Biology, George Mason University, Fairfax, VA 22030, USA
3. Department of Electrical Engineering and Computer Science, Massachusetts Institute of Technology, Cambridge, MA 02139, USA

**\* Correspondence:**
Mostafa Rezapour, PhD
mrezapou@wakehealth.edu

**Keywords:** RNA sequencing, NanoString Technology, Machine Learning, Gene Expression Analysis, Concordance Analysis

**Abstract**

This study evaluates the concordance between RNA sequencing (RNA-Seq) and NanoString technologies for gene expression analysis in non-human primates (NHPs) infected with Ebola virus (EBOV). We performed a detailed comparison of both platforms, demonstrating a strong correlation between them, with Spearman coefficients for 56 out of 62 samples ranging from 0.78 to 0.88, with a mean of 0.83 and a median of 0.85. Bland-Altman analysis further confirmed high consistency, with most measurements falling within 95% confidence limits. A machine learning approach, using the Supervised Magnitude-Altitude Scoring (SMAS) method trained on NanoString data, identified *OAS1* as a key marker for distinguishing RT-qPCR positive from negative samples. Remarkably, when applied to RNA-Seq data, *OAS1* also achieved 100% accuracy in differentiating infected from uninfected samples using logistic regression, demonstrating its robustness across platforms. Further differential expression analysis identified 12 common genes including *ISG15*, *OAS1*, *IFI44*, *IFI27*, *IFIT2*, *IFIT3*, *IFI44L*, *MX1*, *MX2*, *OAS2*, *RSAD2*, and *OASL* which demonstrated the highest levels of statistical significance and biological relevance across both platforms. Gene Ontology (GO) analysis confirmed that these genes are directly involved in key immune and viral infection pathways, reinforcing their importance in EBOV infection. In addition, RNA-Seq uniquely identified genes such as *CASP5*, *USP18*, and *DDX60*, which play key roles in immune regulation and antiviral defense. This finding highlights the broader detection capabilities of RNA-Seq and underscores the complementary strengths of both platforms in providing a comprehensive and accurate assessment of gene expression changes during Ebola virus infection.

## 1. Introduction

RNA sequencing (RNA-Seq) has transformed our capacity to understand gene expression, enabling simultaneous quantification and discovery of transcripts [1]. This high-throughput sequencing technique has been widely adopted, proving essential across various genomic applications including differential gene expression, alternative splicing, and eQTL mapping, among others [1]. For example, Bosworth et al. [2] used RNA sequencing to analyze the cellular mRNA changes in A549 cells infected with the Ebola virus variants Makona and Ecran, focusing on the differential expression of genes. They identified several important genes, including those involved in the inflammatory response, cell proliferation, and leukocyte extravasation, which were differentially expressed and played significant roles in the virus-host interaction. Liu et al. [3] employed RNA sequencing to analyze transcriptome data from peripheral blood samples, which allowed them to identify significant upregulation of interferon signaling and acute phase responses in fatal cases compared to survivors, with key genes like albumin and fibrinogen indicating liver pathology.

On the other hand, NanoString technology complements RNA-Seq with its capability for direct digital quantification of nucleic acids, offering precise and highly multiplexed gene expression analysis without amplification [4]. In our recent application of NanoString, we implemented the Supervised Magnitude-Altitude Scoring (SMAS) methodology, a novel machine learning-based approach, to analyze gene expression data from non-human primates (NHPs) infected with Ebola virus (EBOV) [5]. This approach facilitated the identification of key genes for Ebola infection, enhancing our understanding of host-pathogen interactions. Notably, the use of SMAS led to the discovery of *IFI6* and *IFI27* as key biomarkers, achieving perfect predictive performance with 100% accuracy in differentiating stages of Ebola infection. Additionally, genes such as *MX1*, *OAS1*, and *ISG15* were found to be significantly upregulated, highlighting their key roles in the immune response to EBOV. These findings underscore the effectiveness of NanoString in providing detailed insights into gene expression changes during viral infections, thereby aiding the advancement of diagnostic tools and therapeutic strategies.

Building upon these insights, a question arises: How well do RNA-Seq and NanoString technologies concur in their assessments of gene expression within the context of viral infections? The distinct methodologies, RNA-Seq, with its expansive and detailed transcriptome profiling, and NanoString, known for its precise, targeted quantification, offer potentially complementary views of the same biological phenomena. In our recent study, we analyzed gene expression in human lung organ-tissue equivalents (OTEs) infected with Influenza A virus, Human metapneumovirus, and Parainfluenza virus 3 using both RNA-Seq and NanoString technologies [6]. We demonstrated strong agreement between these platforms in identifying key antiviral genes, thereby validating the reliability of these methods for detailed gene expression analysis in complex infection scenarios [6]. Building on our previous study with OTEs [6], this study launches the first comprehensive evaluation of RNA-Seq and NanoString technologies for gene expression analysis in NHPs infected with the EBOV.

In this study, we aim to ascertain whether these platforms can consistently and accurately capture significant changes in gene expression associated with viral infection. Specifically, we analyze the gene expression data from both technologies to identify and quantify key antiviral genes and investigate their concordance across the two platforms. Given the distinct characteristics of RNA-Seq and NanoString technologies, it is important to employ appropriate statistical methods tailored to their respective data distributions for this study. RNA-Seq, known for capturing a broad array of transcripts, typically exhibits a negative binomial distribution and is prone to overdispersion, a common feature in count data from deep sequencing. To address these challenges, we use Generalized Linear Models with Quasi-Likelihood F-tests and Magnitude-Altitude Scoring (GLMQL-MAS), whose different variants have been validated in previous studies for their efficacy in managing the complexities of RNA-Seq data [7, 8, 9]. This approach ensures robust differential expression analysis, enhancing our ability to detect and quantify the dynamic changes in gene expression associated with EBOV infection across the different platforms.

To assess the concordance between the two platforms, we perform several key analyses. First, we conduct a correlation analysis to compare gene expression profiles across the platforms, using a common set of 584 genes from 62 samples collected over the progression of Ebola virus disease (EVD) in NHPs. The Spearman correlation coefficient [10] is employed to assess the consistency of gene expression patterns across the two methods. Next, we use Bland-Altman analysis [11] to further evaluate the agreement between RNA-Seq and NanoString, examining any systematic biases in gene expression measurements. Finally, we apply machine learning methods to identify key genes from NanoString data that differentiate RT-qPCR positive from negative samples, and evaluate their predictive performance in RNA-Seq data, testing their cross-platform utility. Through these approaches, we aim to confirm the reliability of these technologies for gene expression analysis in complex infection models like EBOV, providing insights into viral pathogenesis and host response.

## 2. Materials and Methods

This study uses RNA-Seq and NanoString gene expression data derived from NHPs infected with the EBOV, as provided and detailed by Speranza et al. [12]. In their study, Speranza et al. [12] focused on enhancing the accuracy of NHP models to more closely reflect human EVD. Their study detailed the exposure of 12 cynomolgus macaques to the EBOV/Makona strain via intranasal routes using a target dose of 100 plaque-forming units (PFU) [12]. Administration methods varied between using a pipette or a mucosal atomization device, leading to diverse symptom onset and disease progression, culminating in four distinct response groups (see **Figure 1**) with an overall fatality rate of 83% [12].

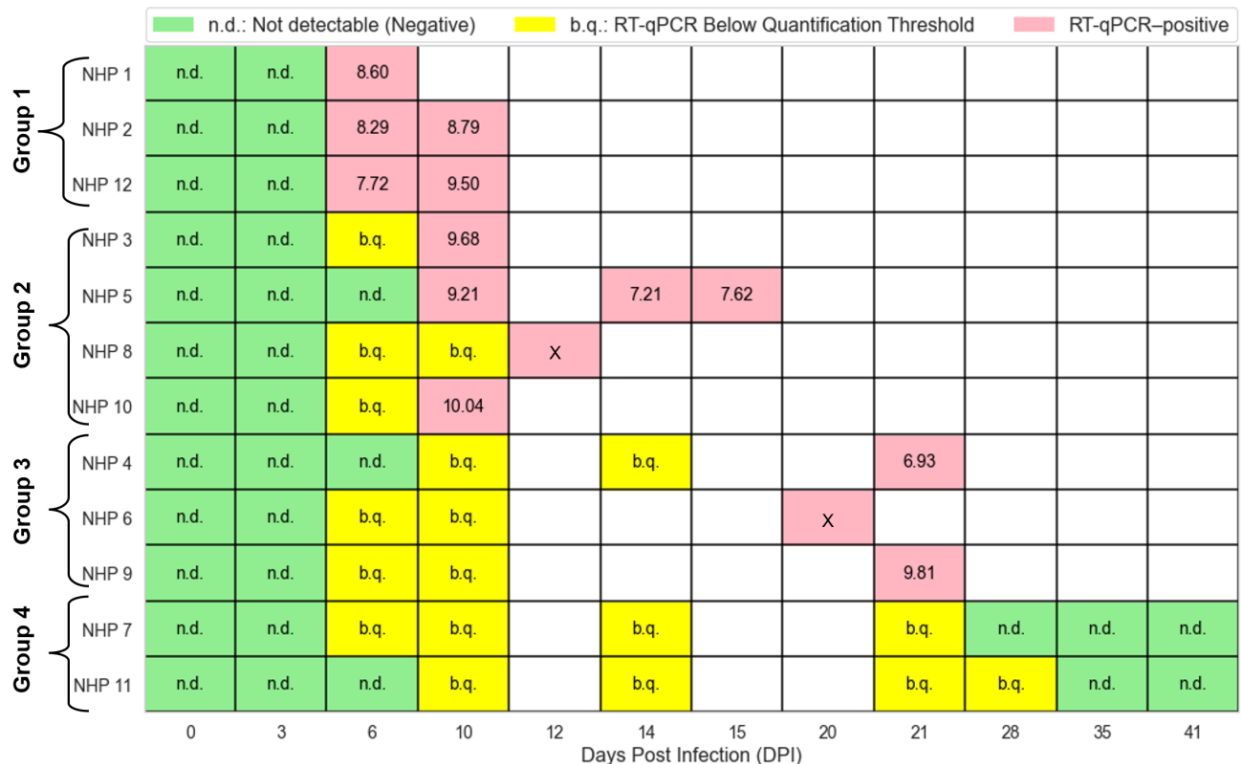

*Figure 1: Timeline of RT-qPCR results for NHPs exposed to EBOV. This figure shows a schedule of test results over various days post-exposure. Green cells indicate non-detectable viral RNA, suggesting negative results, while pink cells denote RT-qPCR positive results with corresponding Genome Equivalent (GE). The numbers displayed in the cells for positive cases represent the GE values. The box marked with an "X" indicates that although the RT-qPCR results are positive, the corresponding NanoString information is unavailable.*

These groups were categorized based on the timing and nature of symptom appearance, onset of viremia, and time to death, ranging from typical EVD courses to those not developing detectable viremia during the study period. Group 1 displayed typical symptoms of EVD with viremia measurable from day 6 and a mean survival time of 10.47 days. In Group 2, viremia was detectable later, between days 10 to 12, and the average survival extended to 13.31 days. Group 3 showed even further delayed signs of the disease, with viremia apparent after day 20 and an average survival time of 21.42 days. In contrast, Group 4 exhibited no detectable viremia throughout the study and survived until the end of the 41-day experiment [12]. In this study, we focus exclusively on NHPs with both RNA-Seq and NanoString data available (see **Figure 1**). For notation purposes, throughout this paper, "NHP-$m$-$n$" refers to NHP #$m$ on day $n$ post-infection.

For the NanoString analysis, 769 specific NHP transcripts were targeted, offering rapid processing and lower RNA quality requirements than RNA-Seq, which makes it particularly effective for Ebola virus research [5, 12]. Standard normalization procedures were followed, involving background adjustments using negative controls and lane variation corrections with internal positive controls [12]. The most stable reference genes for these adjustments were identified using the NormFinder software package [12].

For RNA isolation from whole blood, samples were diluted with molecular biology-grade water and treated with TRIzol LS, followed by purification using the PureLink RNA Mini Kit (Thermo Fisher Scientific) and quality assessment on the Agilent 2200 TapeStation [12]. RNA-seq libraries were prepared using the TruSeq Stranded Total RNA Library Prep Kit (Illumina), with library quality assessed on the TapeStation and quantified via qPCR using the KAPA Complete (Universal) qPCR Kit (Kapa Biosystems) [12]. Sequencing was performed on the Illumina HiSeq 2500 in a paired end 2 × 100-base pair, dual-index format [12]. Post-sequencing, the data were processed by trimming low-quality reads and filtering with the FASTX-Toolkit [12]. Alignment was conducted against the cynomolgus macaque genome using Bowtie2 and Tophat [12]. Throughout this study, we implemented Trimmed Mean of M-values (TMM) normalization [13] on RNA-Seq data for use in machine learning, correlation assessments, and various analyses, extending beyond differential expression analysis.

## 2.1. Correlation Analysis:

In this section, we conducted a correlation analysis to compare the gene expression profiles obtained from RNA-Seq and NanoString technologies. The datasets included a common set of 584 genes for 62 samples extracted from 12 NHPs over the progression of EVD as detailed in **Figure 1**. For the correlation analysis, the first step involved aligning the RNA-Seq and NanoString data by ensuring that the 584 genes were identically ordered in both datasets for all samples. This alignment allowed for accurate pairwise comparisons across the corresponding gene expression vectors from the two platforms.

Each NHP was analyzed individually. For each primate, we constructed two 584-dimensional vectors from their RNA-Seq and NanoString data points. We then calculated the Spearman correlation coefficient [10] for each pair of corresponding vectors. This non-parametric measure was selected because it evaluates the monotonic relationship between the ranked variables, providing insights into the consistency of gene expression patterns observed by the two methods regardless of the absolute expression levels.

## 2.2. Bland-Altman Analysis

In this section, the Bland-Altman analysis [11] was employed to assess the agreement between RNA-Seq and NanoString technologies. The analysis commenced by calculating the mean and difference for each gene expression value across the two platforms for each sample. Specifically, for each paired gene expression value, the mean was determined as the average of the RNA-Seq and NanoString results, while the difference was calculated by subtracting the RNA-Seq value from the NanoString value.

Next, Bland-Altman plots were generated to visually inspect the agreement between the two platforms, plotting the differences against the means for each sample. The mean difference, indicating any systematic bias, and the limits of agreement (mean difference ± 1.96 times the standard deviation of the differences) were also calculated. These limits highlight the range where 95% of the differences are expected to fall. Finally, the proportion of data points lying within and outside these limits was evaluated to quantify the consistency between RNA-Seq and NanoString measurements across samples.

## 2.3. Concordance Assessment through Machine Learning Analysis

In this section, we aim to assess the concordance between NanoString and RNA-Seq platforms by first identifying key genes in NanoString data that differentiate RT-qPCR positive from negative samples using logistic regression, then applying these genes as predictors in RNA-Seq to test their cross-platform utility. In terms of machine learning terminology, NanoString data serves as the training set for key gene identification, and RNA-Seq data is treated as the held-out test set. This approach will allow us to evaluate the efficacy of the selected genes in distinguishing positive from negative samples within the RNA-Seq dataset, testing whether the gene signatures identified in NanoString can be directly applied to RNA-Seq with comparable performance.

As shown in **Figure 1**, there are a total of 12 RT-qPCR positive samples for which both NanoString and RNA-Seq data are available. To effectively measure the performance of the logistic regression model [14] using k-fold stratified cross-validation [15] with the selected genes as predictors, we need to ensure that we have a balanced dataset of positive and negative samples. Therefore, we use the samples from all 12 NHPs on day 0 of infection (DPI=0) as the negative samples for key gene identification.

To identify the genes that are capable of separating positive from negative samples within NanoString and use them as predictors for logistic regression, we apply the Supervised Magnitude-Altitude Scoring (SMAS) [5]. For NanoString data, which often adheres to the assumptions required for parametric t-tests, the SMAS first employs two-sample independent t-tests [16] with Benjamini-Hochberg (BH) correction [17]. For genes that achieve significance according to the BH correction, it ranks them based on their statistical significance and biological relevance using the Magnitude-Altitude Score (MAS). The MAS formula is defined as:

$$\text{MAS}_l = |(\log2\text{FC}_l)|^M |(\log_{10}(p_l^{BH}))|^A,$$

for l= 1, 2, …, s, where $s$ is the number of rejected null hypotheses by BH adjusted method ($p_l^{BH} < \alpha = 0.05$). The hyperparameters M and A, both set to 1 in this study, are chosen to balance the adjusted p-value and the log fold change. Finally, via k-fold stratified cross-validation (k=6) [15], the SMAS employs logistic regression [14] using the top MAS-selected gene as the predictor.

Once we have identified the top MAS-selected genes within NanoString, we will directly use them as predictors for logistic regression to evaluate the performance of the model in separating positive from negative groups within RNA-Seq. Note that since the scale of NanoString is different from that of RNA-Seq, due to differences between the two technologies, we apply k-fold stratified cross-validation within the RNA-Seq to ensure that we fine-tune the parameters of the logistic regression.

Since the viral load of NHPs on day 3 of infection (DPI=3) is also not detectable, we additionally use logistic regression via k-fold stratified cross-validation to see whether the selected MAS genes can differentiate both DPI 0 and 3 (DPI=0 and 3) from positive samples within both NanoString and RNA-Seq. In this way, the selected genes will be tested for new samples that are negative but not part of the MAS gene identification process.

Note that in our previous study [5], where we analyzed 769 genes using NanoString, the MAS ranking system was validated, and the top gene selected by our method differentiated positive from negative samples with 100% accuracy in a held-test set, whereas the best performance by EdgeR [18] or DESeq2 [19] was only 72%, using either p-value or logFC (see Table 2 in [5]).

## 2.4. Concordance Assessment through Differential Expression Analysis

In this section, we aim to conduct a differential expression analysis on both the NanoString and RNA-Seq datasets, beginning with 584 genes common to both. Using the MAS ranking system, we will rank these genes by statistical and biological significance after multiple hypothesis testing through the MAS ranking system. The analysis will then extend to all RNA-Seq genes to identify broader significant patterns. Both datasets will be contrasted against a control group consisting of samples from all 12 NHPs on day 0 of infection. The primary goal is to verify if the top BH-significant genes identified, based on the MAS, demonstrate consistency across different datasets and platforms, ensuring the reliability of gene expression findings and assessing concordance across these technological platforms.

In our analysis of RNA-Seq data, we employ GLMQL-MAS to address the challenges of non-normal data distributions and overdispersion typical of these datasets. Traditional methods like the Student's t-test often fail to effectively manage these complexities [20]. The GLMQL-MAS approach integrates the robustness of Generalized Linear Models (GLMs) [21] with the flexibility of quasi-likelihood estimations [22], making it highly effective in capturing the true biological variations among samples.

We use GLMs to model the relationship between gene expression and experimental conditions, focusing on changes from our day 0 baseline to subsequent time points where viral load is detectable in the 12 RT-qPCR–positive samples. This model enables us to accurately quantify gene expression changes attributable to the infection.

After establishing the model, we conduct Quasi-Likelihood (QL) F-tests to assess the significance of the observed differences in gene expression between the baseline and infected samples. These tests, by adjusting for model dispersion, provide more reliable and robust statistical inferences than traditional methods. They are particularly advantageous in handling the inherent complexities of RNA-seq data.

Using these models and tests, we calculate metrics for each gene, e.g. LogFC and p-value. Additionally, the integration of MAS within this framework allows for prioritizing genes based on both their statistical significance (BH adjusted p-value) and biological impact (LogFC). This dual focus ensures that the genes identified as significant are not only statistically validated but also biologically relevant, enhancing our ability to identify truly important biomarkers.

In this study, we conduct differential expression analysis using three distinct approaches to thoroughly examine gene expression changes:

- First, within the NanoString data, we focus solely on the 584 genes common between NanoString and RNA-Seq datasets. This allows us to directly compare the expression profiles between the two platforms for the same set of genes.

- Secondly, we perform a similar analysis within the RNA-Seq data, again limiting our focus to these 584 common genes. This approach ensures that any findings are directly comparable across both technological platforms.

- The third approach expands the scope of our analysis within the RNA-Seq data to include the full set of available protein-coding genes, totaling 14,328. This broader analysis allows us to capture a wider array of biological changes that may be specific to the RNA-Seq technology and not observable within the smaller common gene set.

Note that in the differential expression analysis of RNA-Seq data, the choice of gene set size, whether it is the 584 common genes or the available full set of 14,328 protein-coding genes, influences the outcomes of the analysis, particularly in terms of LogFC and adjusted p-values. This variance arises from differences in the statistical modeling and adjustment processes inherent in the Generalized Linear Models with Quasi-Likelihood F-tests (GLMQL) and the subsequent BH corrections for multiple testing:

- When analyzing a smaller subset of genes (584 common genes), the statistical power of the model is focused but limited. This limitation can restrict the model's ability to detect smaller yet potentially meaningful expression changes that might be more apparent in a larger dataset (14,328 genes). Conversely, a larger dataset provides a broader basis for the model, potentially capturing more subtle variations in gene expression.

- GLMs fitted to a smaller subset of genes (584 common genes) might yield different parameter estimates, affecting LogFC calculations. These estimates can vary because a smaller number of genes may not provide a complete picture of the underlying biological system, possibly skewing the LogFC.

- The inclusion of more genes in the GLMQL enhances the robustness of the quasi-likelihood estimation. With more genes, the model can better account for inter-gene variability and potential co-expression patterns, which might be obscured in smaller datasets. This modeling can lead to different significance levels and thus different p-values for the gene expression changes.

- BH adjustments are affected by the number of tests (genes) analyzed. In a smaller dataset (584 genes), the correction for multiple comparisons is less stringent than in a larger dataset (14,328 genes). This difference means that for the same raw p-value, the adjusted p-value in a smaller dataset might be smaller (less penalized), suggesting higher statistical significance as compared to the same value in a larger dataset.

## 3. Results

### 3.1. Correlation Analysis:

**Figure 2** presents the results of the correlation analysis between RNA-Seq and NanoString gene expression data in NHPs infected with EBOV. The panel (a) shows the distribution of Spearman correlation coefficients across 62 samples, with a mean of 0.83, a standard deviation of 0.06, and a median of 0.85. The majority of the samples, 56, exhibit strong correlation coefficients between 0.78 and 0.88, indicating a high level of agreement between the two platforms, while a few samples fall below 0.78. The panel (b) displays the time course of Spearman correlation coefficients for each NHP in four distinct groups over the progression of EVD, measured by Days Post Infection (DPI).

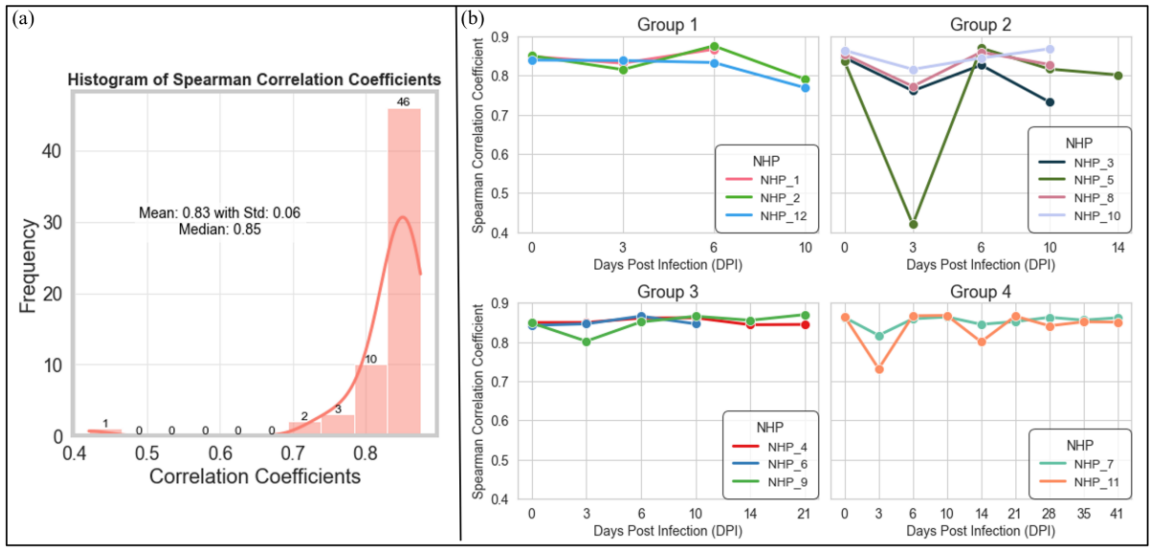

*Figure 2: Histogram and time-course analysis of Spearman correlation coefficients between RNA-Seq and NanoString gene expression data. The panel (a) shows the distribution of Spearman correlation coefficients across 62 samples, with a mean of 0.83, a median of 0.85, and most values concentrated between 0.8 and 0.9. The panel (b) illustrates the Spearman correlation coefficients over Days Post Infection (DPI) for four distinct groups of NHPs, with each line representing an individual NHP.*

## 3.2. Bland-Altman Analysis

**Figure 3** illustrates the agreement between RNA-Seq and NanoString platforms based on Bland-Altman analysis. The panel (a) presents Bland-Altman plots for NHP-1 and NHP-2 on DPI 0, showing the differences between the two platforms plotted against the mean of their measurements. The plots also display the limits of agreement (LOA). For NHP-1-0, 97.95% of the measurements fall within the 95% confidence limits, with a mean difference of 132.28, while 2.05% are outside the limits. For NHP-2-0, 98.46% of the measurements are within the limits, with a mean difference of 128.08 and 1.54% outside the limits, demonstrating a high degree of agreement between the two platforms.

The panel (b) shows the percentage of measurements within the limits of agreement across all samples. The percentage ranges from 97.26% to 99.66%, indicating that the vast majority of gene expression measurements consistently fall within the 95% confidence limits for most samples. Although some samples, like NHP-4-14 and NHP-9-21, exhibit larger mean differences (527.83 and 896.51, respectively), while others like NHP-2-6 show smaller differences (-31.86), the data generally remain within the limits of agreement, demonstrating the robustness of the two platforms in capturing gene expression changes across samples.

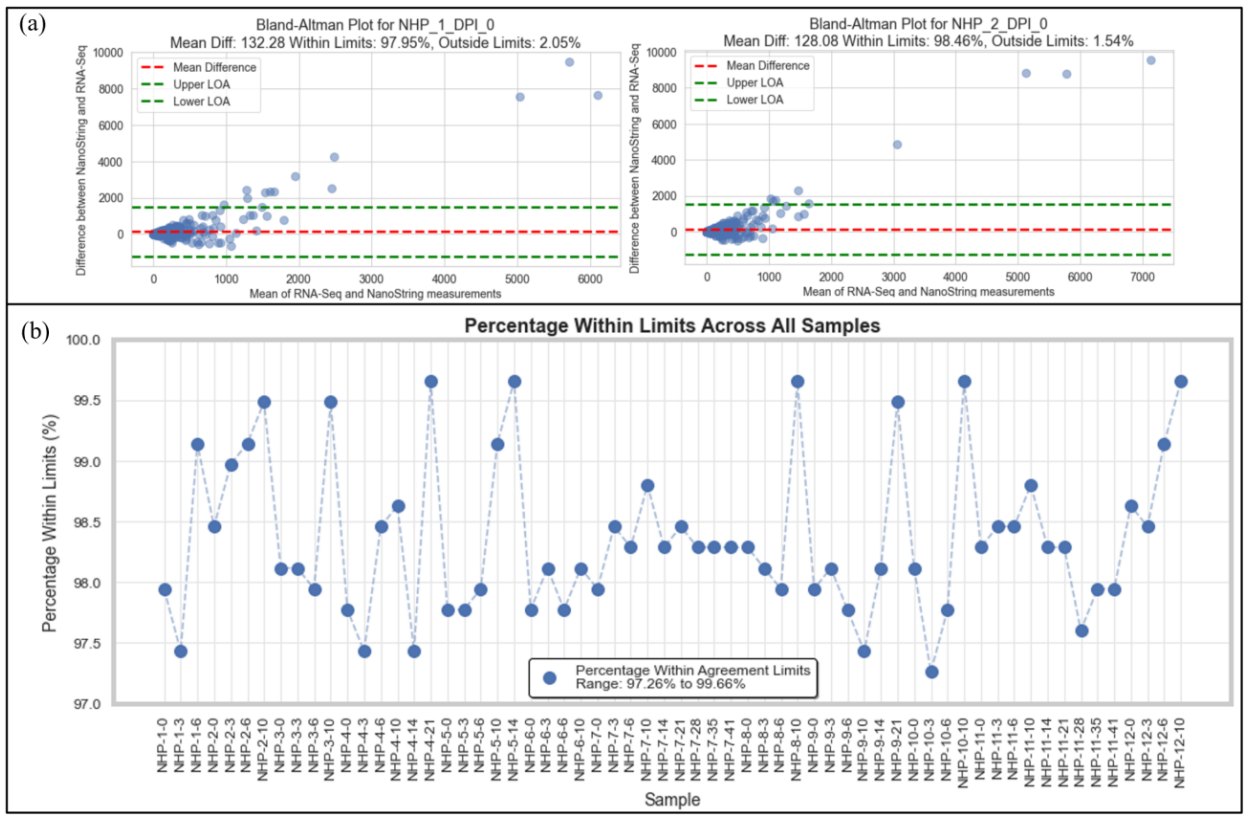

*Figure 3: Bland-Altman analysis comparing RNA-Seq and NanoString platforms. (a) Bland-Altman plots for NHP_1 and NHP_2 on DPI 0, showing the mean difference and limits of agreement between the platforms. (b) Percentage of measurements within the 95% confidence limits across all samples, ranging from 97.26% to 99.66%. "NHP-m-n" refers to NHP #m on day n post-infection.*

## 3.3. Concordance Assessment through Machine Learning Analysis

**Figure 4** illustrates the top-10 MAS-selected genes within NanoString data, using only 584 common genes, after contrasting the positive group against the negative group of NHPs on Day 0 post infection (DPI=0).

Since we have only 12 positive and 12 negative samples in total, to avoid any overfitting, we used only the top gene as the single predictor. Using the top MAS-selected gene, *OAS1*, we applied logistic regression via 6-fold stratified cross-validation to predict whether a sample is from Day 0 (DPI=0) or a positive sample within both NanoString and RNA-Seq data. It turns out that the single gene, *OAS1*, is capable of separating all positive samples from negative samples on day 0 post-infection with 100% accuracy (see **Table 1** and **Figure 5**).

We then added samples from Day 3 (DPI=3) to the negative group to see whether *OAS1* could differentiate NHPs on DPI 0 and 3 from positive samples. It turns out that OAS1 is indeed capable of this differentiation, achieving an average accuracy of 100% (see **Table 1** and **Figure 6**).

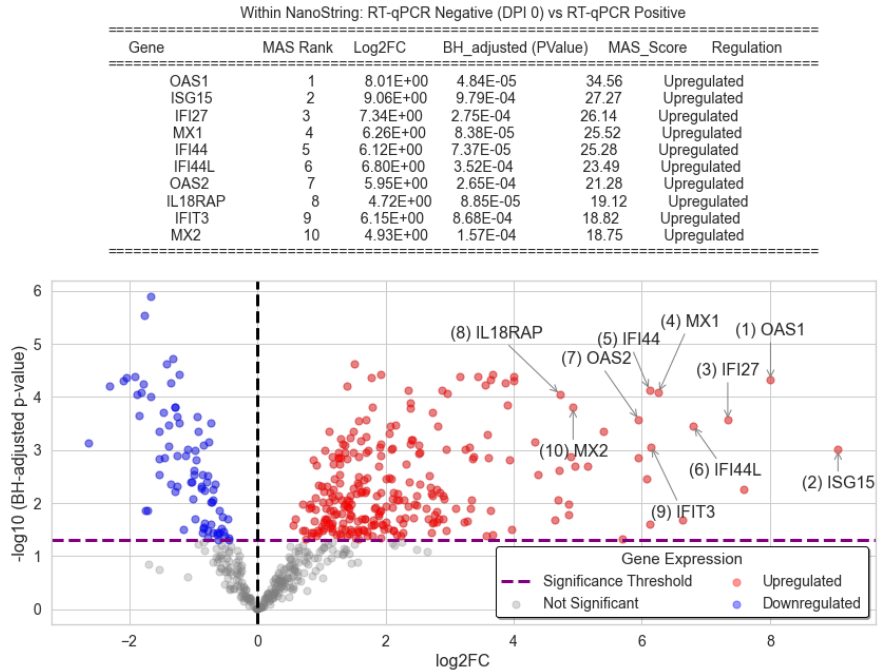

*Figure 4:* Top-10 MAS-selected genes from NanoString analysis contrasting positive and negative NHP groups on day 0 (DPI=0), using 584 common genes.

*Table 1.* Performance comparison of logistic regression model using top genes selected via 6-fold stratified cross validation by MAS within NanoString and RNA-Seq.

| Data | Predictor | Negative Group: DPI | Average AUC | Average Accuracy % | Average Precision | Average Recall |
|---|---|---|---|---|---|---|
| NanoString (MAS training) | *OAS1* | 0 | 1.00 ± 0.00 | 100.00 ± 0.00 | 1.00 ± 0.00 | 1.00 ± 0.00 |
| RNA-Seq (MAS held-out test) | *OAS1* | 0 | 1.00 ± 0.00 | 100.00 ± 0.00 | 1.00 ± 0.00 | 1.00 ± 0.00 |
| NanoString (MAS training) | *OAS1* | 0 and 3 | 1.00 ± 0.00 | 100.00 ± 0.00 | 1.00 ± 0.00 | 1.00 ± 0.00 |
| RNA-Seq (MAS held-out test) | *OAS1* | 0 and 3 | 1.00 ± 0.00 | 100.00 ± 0.00 | 1.00 ± 0.00 | 1.00 ± 0.00 |

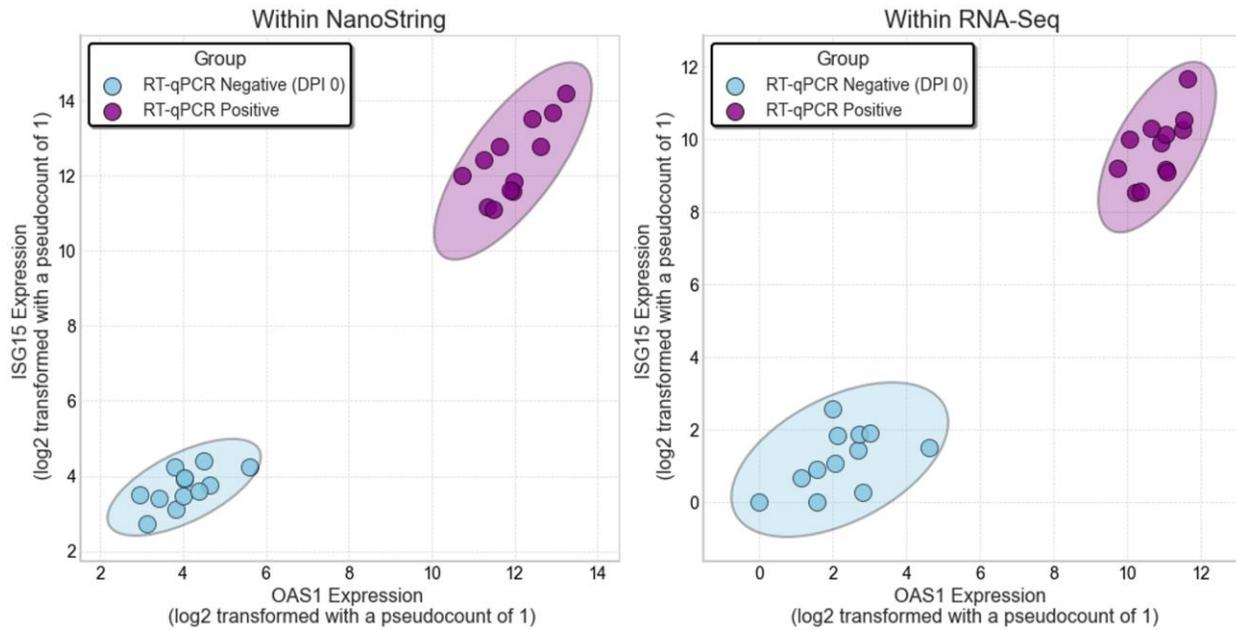

*Figure 5:* *Performance of the OAS1 gene in differentiating Day 0 (DPI=0) negative samples from positive samples within both NanoString and RNA-Seq data.*

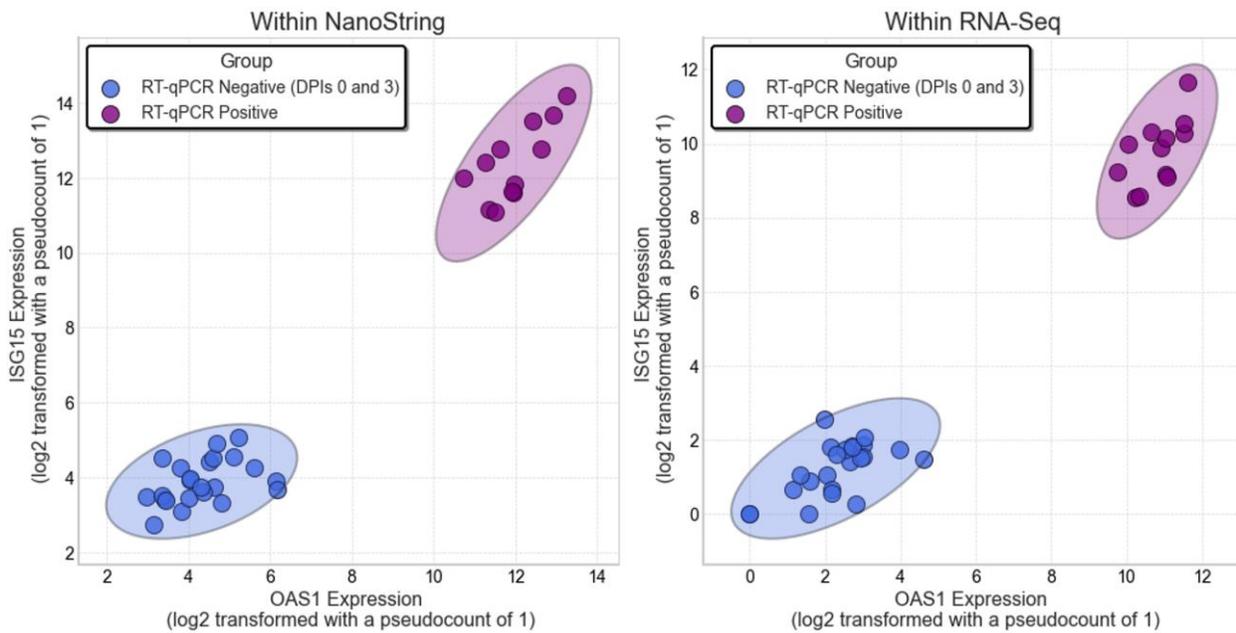

*Figure 6:* *Performance of the OAS1 gene in differentiating Days 0 and 3 (DPI=0 and 3) negative samples from positive samples within both NanoString and RNA-Seq data.*

### 3.4. Concordance Assessment through Differential Expression Analysis

**Figure 7** illustrates the top 20 MAS-selected genes within three datasets: RNA-Seq (Common genes), NanoString (Common genes), and RNA-Seq (Full genes), where "Common genes" indicates that only 584 common genes between the two platforms were used during the analysis. According to **Figure 7**, genes unique to the RNA-Seq (Common genes) dataset within top 20 MAS selected genes include *AHR*, *IFIT5*, and *TCF7*. The NanoString (Common genes) dataset showed exclusivity for *IL18RAP*, *S100A8*, *DHX58*,

*SIGLEC1*, *S100A9*, *DYSF*, and *TWIST2*. In contrast, the RNA-Seq (Full genes) dataset exclusively featured *CASP5*, *USP18*, *DDX60*, and *PLA2G4C*.

Shared gene signatures were also observed, with *ISG15*, *OAS1*, *IFI44L*, *IFIT2*, *OAS2*, *MX1*, *IFIT3*, *IFI44*, *OASL*, *IFI27*, *RSAD2*, *MX2*, and *CCL8* common between RNA-Seq (Common genes) and NanoString. Similarly, RNA-Seq (Common genes) and RNA-Seq (Full genes) shared *ISG15*, *OAS1*, *IFI44L*, *IFIT2*, *OAS2*, *MX1*, *IFIT3*, *IFI44*, *OAS3*, *OASL*, *IFI27*, *RSAD2*, *DDX58*, *FCGR1A*, *MX2*, and *TLR3*. Genes shared between NanoString and RNA-Seq (Full genes) included *OAS1*, *ISG15*, *IFI27*, *MX1*, *IFI44*, *IFI44L*, *OAS2*, *IFIT3*, *MX2*, *OASL*, *IFIT2*, and *RSAD2*. Notably, genes common across all datasets, emphasizing their robust cross-platform consistency, were *IFIT2*, *IFI44*, *OASL*, *IFI27*, *IFIT3*, *IFI44L*, *MX1*, *OAS1*, *MX2*, *OAS2*, *RSAD2*, and *ISG15*.

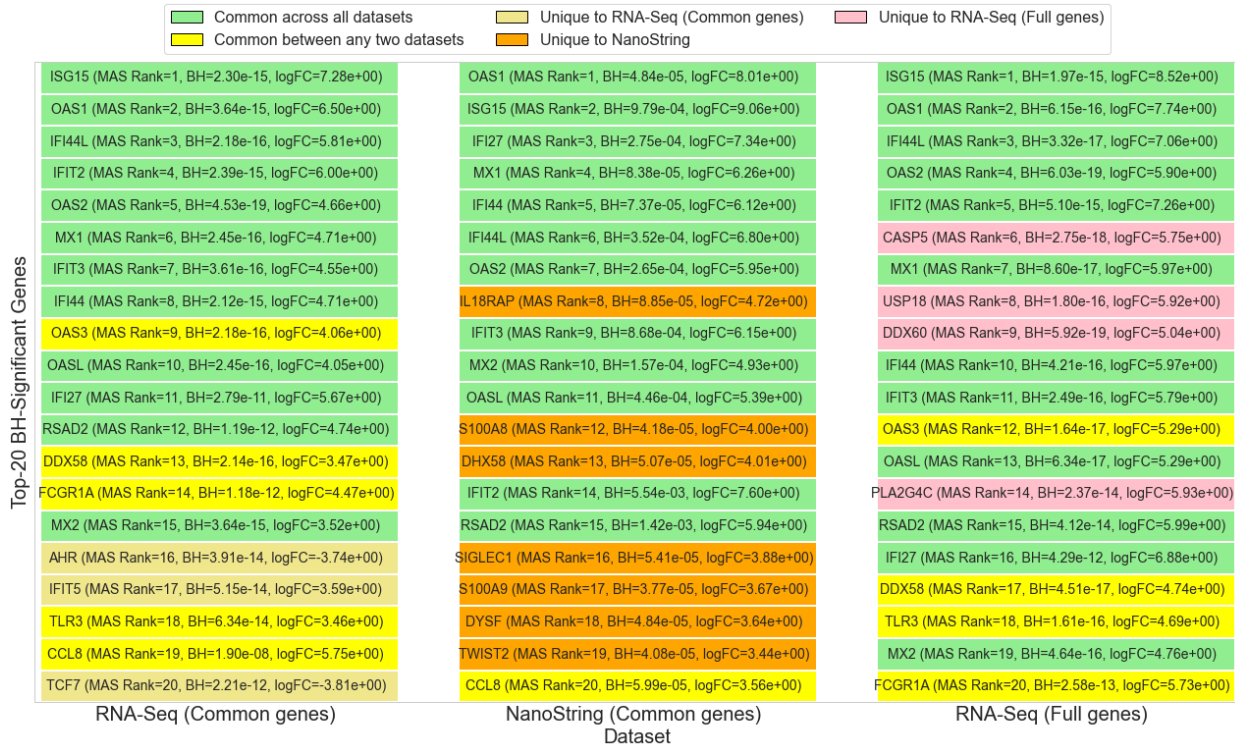

*Figure 7:* Visualization of the top-20 MAS-selected genes within three datasets: RNA-Seq (Common genes), NanoString (Common genes), and RNA-Seq (Full genes). The graph displays genes that are unique to each dataset as well as those shared among them. Each gene is color-coded to indicate its uniqueness or commonality.

We proceed to test the effectiveness of these 12 common genes, *IFIT2*, *IFI44*, *OASL*, *IFI27*, *IFIT3*, *IFI44L*, *MX1*, *OAS1*, *MX2*, *OAS2*, *RSAD2*, and *ISG15*, in clustering RNA-Seq and NanoString samples. Hierarchical clustering [23] is performed using Ward's method [24] with Euclidean distance as the metric. This analysis aims to validate whether these genes can effectively group the samples based on similarities in their expression profiles across platforms, confirming their discriminative power and relevance in broader genomic studies. **Figures 8** and **9** illustrate the hierarchical clustering within RNA-Seq and NanoString datasets, respectively.

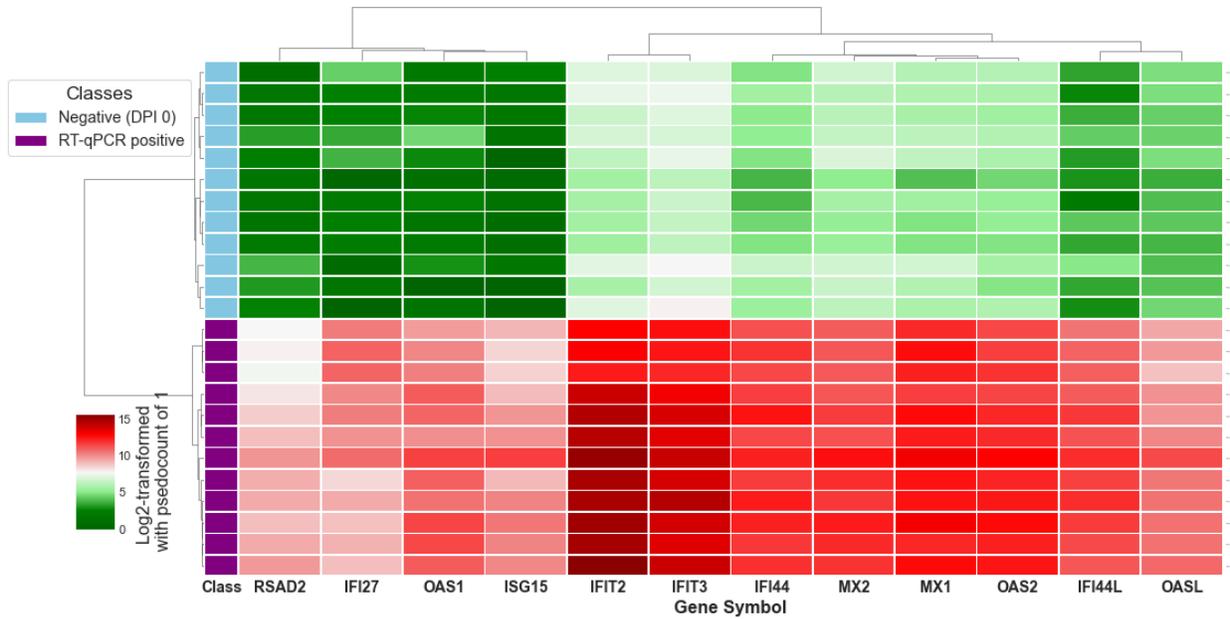

*Figure 8:* *Hierarchical clustering of RNA-Seq samples using Ward's method and Euclidean distance. This visualization demonstrates the grouping of samples based on the expression profiles of 12 common genes, highlighting their discriminative power across RNA-Seq datasets.*

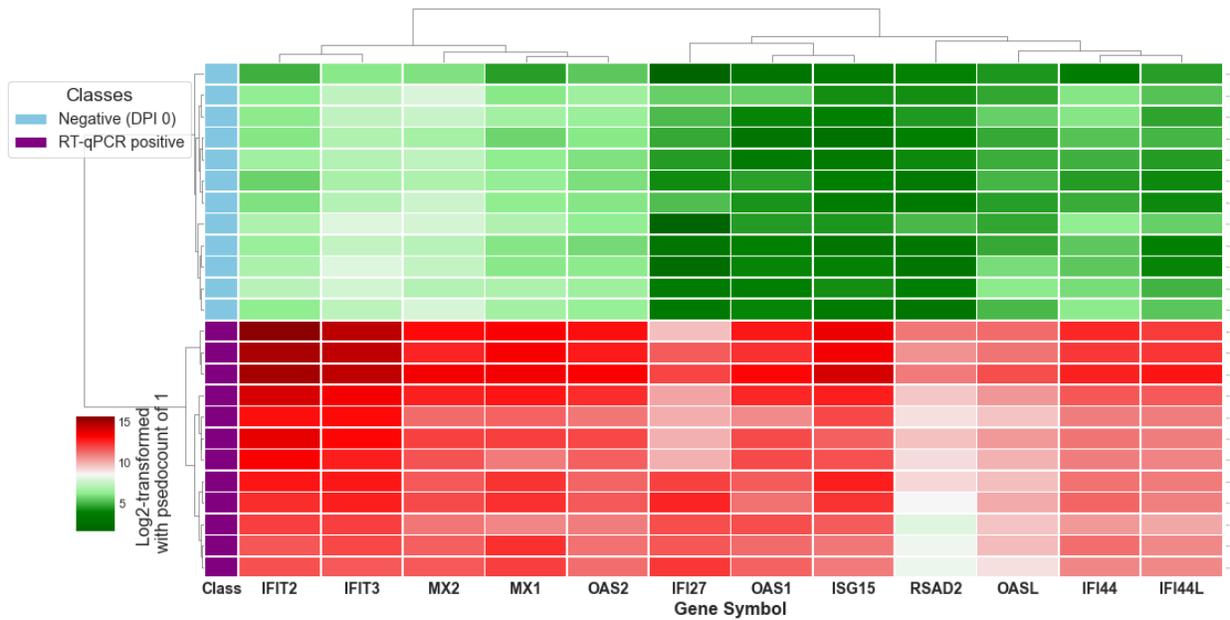

*Figure 9:* *Hierarchical clustering of NanoString samples using Ward's method and Euclidean distance. This figure illustrates how the 12 common genes cluster NanoString samples, validating the consistency and relevance of these genes in gene expression analysis across different platforms.*

To systematically identify the biological processes associated with these 12 common genes, we employed the MyGeneInfo API, an accessible resource for gene-related data provided by the MyGene.info web service. We used version 3 of the API, which offers extensive data access via Python through the mygene Python package. This package facilitates queries against gene symbols and retrieves data about associated Gene Ontology (GO) terms, specifically focusing on biological processes. The comprehensive list of biological processes identified for each gene, as retrieved from the MyGeneInfo API, is detailed in **Table 2.**

*Table 2: This table summarizes the Gene Ontology (GO) biological process terms for 12 genes implicated in the immune response to viral infections, as identified through the MyGeneInfo API.*

| Gene | Biological Processes |
|---|---|
| OAS1 | Glucose metabolic process, response to virus, positive regulation of interferon-beta production, positive regulation of tumor necrosis factor production, toll-like receptor 3 signaling pathway, toll-like receptor 4 signaling pathway, cellular response to interferon-alpha, cellular response to interferon-beta, glucose homeostasis, defense response to bacterium, surfactant homeostasis, negative regulation of viral genome replication, protein complex oligomerization, defense response to virus, type I interferon-mediated signaling pathway, negative regulation of type I interferon-mediated signaling pathway, regulation of ribonuclease activity, interleukin-27-mediated signaling pathway, positive regulation of monocyte chemotactic protein-1 production, negative regulation of IP-10 production, cellular response to virus, antiviral innate immune response, positive regulation of cellular respiration, negative regulation of chemokine (C-X-C motif) ligand 2 production. |
| IFIT2 | Apoptotic mitochondrial changes, response to viruses, negative regulation of protein binding, positive regulation of apoptotic processes, defense against viruses, antiviral innate immune response. |
| IFI44 | Immune response, response to viruses, response to bacterium |
| IFI27 | Negative regulation of transcription by RNA polymerase II, release of cytochrome c from mitochondria, apoptotic processes, proteasome-mediated ubiquitin-dependent protein catabolic process, modulation by host of viral genome replication, innate immune responses, regulation of protein export from nucleus, various signaling pathways linked to the immune system. |
| IFIT3 | Negative regulation of cell population proliferation, response to viruses, negative regulation of apoptotic process, defense against viruses, antiviral innate immune response. |
| IFI44L | Immune response, defense against viruses. |
| MX1 | Apoptotic process, defense response, signal transduction, response to type I interferon, negative regulation of viral genome replication, innate immune responses, defense against viruses, interleukin-27-mediated signaling pathways. |
| MX2 | Defense response, protein transport, response to interferon-alpha, innate immune response, regulation of nucleocytoplasmic transport, mRNA transport, regulation of the cell cycle. |
| OAS2 | Nucleobase-containing compound metabolic process, RNA catabolic process, response to viruses, positive regulation of interferon-beta production, defense responses to bacteria, negative regulation of viral genome replication, type I interferon-mediated signaling pathways. |
| RSAD2 | Response to viruses, positive regulation of toll-like receptor signaling pathways, T cell activation and differentiation, negative regulation of viral genome replication, innate immune responses, negative regulation of protein secretion, positive regulation of immune responses, defense responses to viruses. |
| OASL | Response to virus, negative regulation of viral genome replication, innate immune response, defense response to virus, type I interferon-mediated signaling pathway, interleukin-27-mediated signaling pathway, antiviral innate immune response, positive regulation of RIG-I signaling pathway. |
| ISG15 | Integrin-mediated signaling pathway, response to viruses, protein ubiquitination, modification-dependent protein catabolic process, positive regulation of bone mineralization, negative regulation of protein ubiquitination, ISG15-protein conjugation, regulation of type II interferon production, positive regulation of interferon-beta production, interleukin |

| | production, response to type I interferon, defense against bacteria, innate immune responses, various other processes linked to cellular functions and immune modulation. |

Note that there are some genes in RNA-Seq (Full genes) that have not appeared in the NanoString, such as *CASP5*, *USP18*, and *DDX60*. **Figure 10** illustrates a 3-dimensional visualization of RNA-Seq samples using these three genes as coordinates after a log2 transformation. Employing MyGeneInfo API, we found that *CASP5* is involved in the positive regulation of the inflammatory response. *USP18* participates in the negative regulation of type I interferon-mediated signaling pathways and the antiviral innate immune response. *DDX60* is associated with the response to viruses, the innate immune response, defense against viruses, and the positive regulation of the MDA-5 and RIG-I signaling pathways.

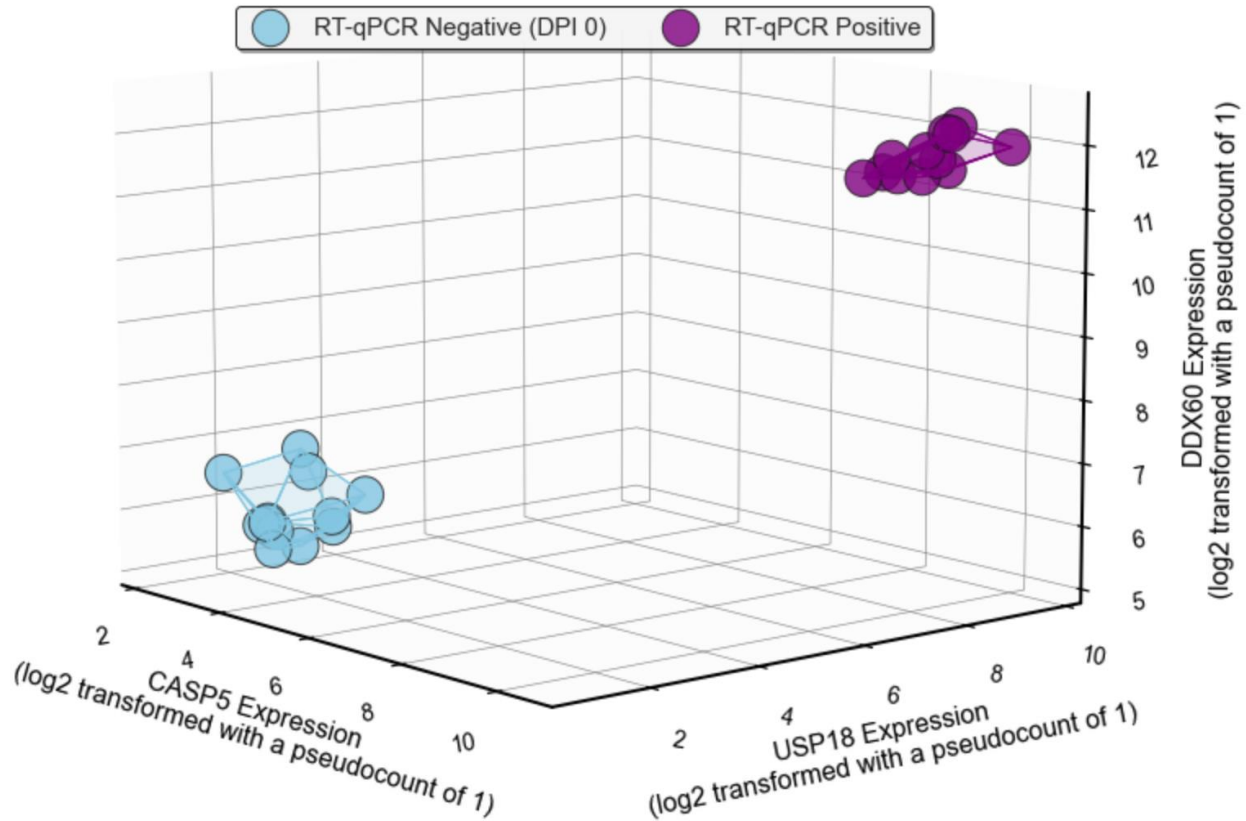

*Figure 10: Three-dimensional visualization of RNA-Seq samples using CASP5, USP18, and DDX60 as coordinates following log2 transformation with a pseudocount of 1.*

## 4. Discussion

### 4.1. Correlation Analysis:

The correlation analysis illustrated in **Figure 2** indicates a strong agreement between the RNA-Seq and NanoString gene expression data for NHPs exposed to the EBOV. From the panel (a) of **Figure 2**, the notable concentration of Spearman correlation coefficients in the range of 0.78 to 0.88, with 56 out of 62 samples falling within this interval, strongly suggests that both technologies provide highly consistent measurements of gene expression. This high degree of correlation, supported by a mean of 0.83 and a median of 0.85, confirms that both RNA-Seq and NanoString are effective in capturing the biological responses of NHPs to the virus, thus validating the use of these platforms in parallel for comprehensive gene expression analysis in virology research.

In the panel (b) of the figure, the Spearman correlation coefficients between RNA-Seq and NanoString gene expression data are shown for four distinct groups of NHPs over the course of EVD. Group 1 exhibits consistent correlations throughout the infection period, with values ranging from 0.8 to 0.9, indicating minimal fluctuations in agreement between the two platforms. Group 2 shows mostly stable correlations, but NHP-5 displays a significant dip at DPI 3, suggesting variability in gene expression consistency at this stage, before returning to levels similar to other NHPs in the group. Group 3 demonstrates strong and stable correlations across all NHPs, with very little variation throughout the infection period. In Group 4, correlations remain stable overall, indicating general agreement between RNA-Seq and NanoString with minor fluctuations.

### 4.2. Bland-Altman Analysis

The Bland-Altman analysis shown in **Figure 3** demonstrates a strong agreement between RNA-Seq and NanoString technologies, with the majority of measurements falling within the 95% confidence limits across all samples. Despite some variations in the mean differences between the two platforms for certain samples, the consistently high percentage of measurements within the limits suggests that both technologies provide reliable and comparable gene expression data. Overall, the Bland-Altman analysis confirms the robustness of these platforms in measuring gene expression with minimal bias, making them effective tools for this study.

### 4.3. Concordance Assessment through Machine Learning Analysis

The machine learning analysis, employing the SMAS, effectively used the strengths of both NanoString and RNA-Seq technologies to enhance our understanding of gene expression in NHPs infected with the EBOV. Initially, we identified key antiviral genes using the precise quantification capabilities of NanoString. *OAS1* was selected as a primary marker due to its significant ability to differentiate between positive and negative samples, as determined by the MAS ranking system (**Figure 4**). Employing logistic regression with six-fold stratified cross-validation, we observed that *OAS1* accurately differentiated between negative (DPI 0) and positive samples in the NanoString dataset, achieving 100% accuracy. Subsequently, OAS1 was used as the sole predictor in logistic regression within RNA-Seq to determine whether the top MAS-selected gene from NanoString could extend its capability to differentiate between positive and negative samples. Indeed, OAS1 demonstrated this capability (**Table 1** and **Figure 5**).

Furthermore, by applying the MAS-selected gene, OAS1, directly within the RNA-Seq analysis using the same stratified cross-validation method, we also achieved 100% accuracy in differentiating extended negative group (DPI 0 and DPI 3) from positive samples (**Table 1** and **Figure 6**). The successful application of a NanoString-identified gene in RNA-Seq classification underscores the complementary potential of these platforms. It demonstrates how findings from one technology can be substantiated and extended in another, offering a methodological blueprint for future studies that aim to leverage multiple genomic technologies.

### 4.4. Concordance Assessment through Differential Expression Analysis

From **Figure 7**, we observed a strong concordance between the NanoString and RNA-Seq platforms, as 12 out of the top 20 genes were common across all three datasets. These shared genes, including *ISG15*, *OAS1*, *IFI44*, and *RSAD2*, represent key antiviral responses that were consistently identified across both technologies. This overlap suggests that despite their technological differences, both RNA-Seq and NanoString are effective in capturing key gene expression changes during Ebola virus infection.

The hierarchical clustering shown in **Figures 8** and **9** highlights the effectiveness of the 12 common genes in differentiating samples based on their gene expression profiles. These genes successfully grouped RNA-Seq (**Figure 8**) and NanoString (**Figure 9**) samples into distinct clusters, indicating their strong discriminative power. In both datasets, the clustering distinctly separates RT-qPCR positive samples from negative ones (DPI 0). This clear division suggests that the selected genes, *IFIT2*, *IFI44*, *OASL*, *IFI27*, *IFIT3*, *IFI44L*, *MX1*, *OAS1*, *MX2*, *OAS2*, *RSAD2*, and *ISG15*, are highly responsive to viral infection, consistently reflecting changes in gene expression across both platforms. The comparable clustering patterns across RNA-Seq and NanoString datasets reinforce the concordance between the two technologies, validating these genes as reliable biomarkers for distinguishing between infected and uninfected samples in Ebola virus infection studies.

**Table 2** provides a comprehensive overview of the biological processes associated with the 12 common genes that were consistently identified across both the RNA-Seq and NanoString platforms. The data retrieved through the MyGeneInfo API highlights the involvement of these genes in key immune responses, particularly antiviral defenses, which include processes like the innate immune response, regulation of interferon production, and negative regulation of viral genome replication. This overlap not only underscores the reliability of RNA-Seq and NanoString in capturing essential gene functions but also confirms the biological significance of the identified genes.

Using *CASP5*, *USP18*, and *DDX60*, the RT-qPCR negative (DPI 0) samples and the RT-qPCR positive samples are distinctly separated, as shown in **Figure 10**. These genes, which did not appear in the NanoString, highlight the broader gene detection capabilities of RNA-Seq. Their biological processes, previously discussed in Section 3, reflect critical roles in immune regulation and antiviral defense mechanisms. The ability of these genes to clearly differentiate infected from uninfected samples underscores the importance of RNA-Seq in identifying additional biologically significant markers, which may be overlooked by more targeted platforms like NanoString. This further illustrates the complementary nature of the two platforms in gene expression analysis.

## 5. Conclusions

This study offers a comprehensive evaluation of the concordance between RNA-Seq and NanoString technologies for gene expression analysis NHPs infected with EBOV. Our results demonstrate that both platforms provide highly consistent gene expression measurements, as evidenced by strong Spearman correlation coefficients and Bland-Altman analyses, confirming their reliability in capturing complex biological responses in viral infections.

Machine learning analysis using the SMAS method, trained on NanoString data, identified *OAS1* as a key marker capable of distinguishing RT-qPCR positive from negative samples. When applied to RNA-Seq data, *OAS1* achieved 100% accuracy using logistic regression, underscoring its robustness and cross-platform utility. This finding highlights the potential of NanoString-identified genes to be validated and extended in RNA-Seq datasets, reinforcing the complementary nature of these technologies.

Additionally, differential expression analysis identified 12 common genes including *ISG15*, *OAS1*, *IFI44*, *IFI27*, *IFIT2*, *IFIT3*, *IFI44L*, *MX1*, *MX2*, *OAS2*, *RSAD2*, and *OASL* that exhibited the greatest statistical significance and biological relevance across both platforms. These genes are primarily associated with antiviral immune responses and were shown to reliably differentiate infected from uninfected samples through hierarchical clustering. Gene Ontology analysis further confirmed their involvement in immune pathways, reinforcing their potential as key biomarkers for Ebola virus infection.

RNA-Seq also uniquely identified genes such as *CASP5*, *USP18*, and *DDX60*, which are involved in immune regulation and antiviral defense mechanisms. These genes, not captured in NanoString, emphasize the broader detection capabilities of RNA-Seq, making it particularly useful for discovering additional biologically relevant markers in complex infection scenarios.

In conclusion, this study demonstrates that RNA-Seq and NanoString technologies are both powerful tools for gene expression analysis in EBOV-infected NHPs. Their complementary strengths, NanoString's precision in quantification and RNA-Seq's broader gene detection, provide a comprehensive understanding of gene expression dynamics. This cross-platform approach enhances the reliability of identified biomarkers and offers valuable insights for future research on viral infections, therapeutic targets, and disease monitoring.

## 6. Limitations of the Study

One limitation of this study is the reliance on a single animal model (non-human primates) for investigating Ebola virus infection, which may limit the generalizability of the findings to other species, including humans. Additionally, the sample size was relatively small, which could affect the robustness of the statistical analyses and the ability to detect more subtle gene expression differences. Finally, while the study demonstrates concordance between the two technologies, the validation of identified biomarkers in additional cohorts or models is necessary to confirm their broader applicability in Ebola virus research.

## 7. Acknowledgement

## 8. Author Contribution

Conceptualization, M.R., A.N., W.H.M., and M.N.G.; methodology, M.R., A.N., and M.N.G.; software, M.R.; validation, M.R., A.N., and M.N.G.; formal analysis, M.R., A.N., and M.N.G.; investigation, M.R., A.N., and M.N.G.; resources, A.N. and M.N.G.; data curation, M.R., and W.H.M.; writing-original draft preparation, M.R., A.N., and M.N.G.; writing-review and editing, M.R., A.N., W.H.M., and M.N.G.; visualization, M.R.; supervision, A.N. and M.N.G.; project administration, A.N. and M.N.G.; funding acquisition, A.N. All authors have read and agreed to the published version of the manuscript.

## 9. Data Availability

The data supporting the findings of this study, originally generated by Speranza et al. [12] are openly available in the Gene Expression Omnibus (GEO) repository hosted by the National Center for Biotechnology Information (NCBI). The normalized NanoString count data can be found in table S3 at: https://www.science.org/doi/10.1126/scitranslmed.aaq1016. The RNA-seq data can be accessed through Gene Expression Omnibus accession number GSE103825.